\pdfoutput=1
\documentclass[%
 aip,
 amsmath,amssymb,
twocolumn,
longbibliography,
 reprint,%
]{revtex4-1}

\usepackage{hyperref}
\hypersetup{colorlinks,breaklinks,linkcolor=Blue,urlcolor=Blue,anchorcolor=Blue,citecolor=Blue}
\usepackage{color}
\definecolor{DarkBlue}{rgb}{0.0,0.08,0.45}
\definecolor{Blue}{rgb}{0.0,0.0,1.0}
\definecolor{Red}{rgb}{1.0,0.0,0.0}
\definecolor{RedOrange}{rgb}{0.9,0.0,0.2}
\definecolor{dgreen}{RGB}{0,128,0}
\definecolor{dgray}{gray}{0.3}

\newcommand*{\citen}{}
\DeclareRobustCommand*{\citen}[1]{%
  \begingroup
    \romannumeral-`\x 
    \setcitestyle{numbers}%
    \cite{#1}%
  \endgroup
}

\newcommand{\mybra}[1]{\langle#1|}
\newcommand{\myket}[1]{|#1\rangle}

\newcommand{\nbd}{\nobreakdash}
\newcommand{\MP}{M{\o}ller\nbd-Plesset}

\usepackage{graphicx}
\usepackage{dcolumn}
\usepackage{bm}

\usepackage[utf8]{inputenc}
\usepackage[T1]{fontenc}
\usepackage{mathptmx}
\usepackage{etoolbox}

\usepackage{comment}

\newcolumntype{.}{D{.}{.}{-1}}
\newcommand{\mc}[3]{\multicolumn{#1}{#2}{#3}}

\usepackage{pifont}
\newcommand{\cmark}{\ding{51}}
\newcommand{\xmark}{\ding{55}}

\newcommand{\fig}[2]{\scalebox{#1}{\includegraphics{#2}}}

\makeatletter
\def\@email#1#2{%
 \endgroup
 \patchcmd{\titleblock@produce}
  {\frontmatter@RRAPformat}
  {\frontmatter@RRAPformat{\produce@RRAP{*#1\href{mailto:#2}{#2}}}\frontmatter@RRAPformat}
  {}{}
}%
\makeatother
\begin{document}

\preprint{AIP/123-QED}

\title{
Repartitioned Brillouin-Wigner Perturbation Theory with a Size-Consistent Second-Order Correlation Energy
}
\author{Kevin Carter-Fenk}
\email{carter-fenk@berkeley.edu}
\affiliation{ 
Kenneth S. Pitzer Center for Theoretical Chemistry, Department of Chemistry, University of California, Berkeley, CA 94720, USA.
}
\author{Martin Head-Gordon}%
 \email{mhg@cchem.berkeley.edu}
 \affiliation{ 
Kenneth S. Pitzer Center for Theoretical Chemistry, Department of Chemistry, University of California, Berkeley, CA 94720, USA.
}
\affiliation{
Chemical Sciences Division, Lawrence Berkeley National Laboratory, Berkeley, CA 94720, USA
}

\date{\today}

\begin{abstract}
Second-order M{\o}ller-Plesset perturbation theory (MP2)
often breaks down catastrophically in small-gap systems,
leaving much to be desired in its performance for myriad chemical applications
such as noncovalent interactions, thermochemistry, and
dative bonding in transition metal complexes.
This divergence problem has reignited interest
in Brillouin-Wigner perturbation theory (BWPT), which
is regular at all orders
but lacks size-consistency and extensivity,
severely limiting its application to chemistry.
In this work, we propose an alternative partitioning of the
Hamiltonian that leads to a regular BWPT perturbation
series that, through second order,
is size-extensive, size-consistent (provided its Hartree-Fock reference is also), and
orbital invariant.
Our second-order size-consistent
Brillouin-Wigner (BW-s2)
approach is capable of describing the exact dissociation limit of
H$_2$ in a minimal basis set regardless of the spin-polarization of
the reference orbitals. 
More broadly, we find that BW-s2 offers improvements relative to MP2 for
covalent bond breaking,
noncovalent interaction energies, and
metal\slash organic reaction energies,
while rivaling coupled-cluster with single and double
substitutions (CCSD) for thermochemical properties.
\end{abstract}

\maketitle

\section{\label{sec:Intro} Introduction}

The oldest and most tractable wave function approach that captures
electron correlation from first principles is
second-order \MP\ perturbation theory (MP2).
While the ${\mathcal O}(N^5)$
asymptotic scaling of MP2\cite{Cre11}
does not compete with the ${\mathcal O}(N^3)$ scaling of
density functional theory (DFT),
MP2 is immune to many of the nonphysical problems
that manifest in DFT such as self-interaction error,
which can obfuscate the underlying physics of chemical systems
by artificially delocalizing charge density.\cite{ZhaYan98b,MorCohYan06,MorCohYan08}
The {\em ab initio}, and therefore self-interaction-free, correlation
offered by MP2 has led to its incorporation into double-hybrid
density functionals,
which combine MP2 with DFT
exchange-correlation.\cite{ShaTouSav11,SanAda13,GoeGri14,BreCioSan16,KalTou18,MarSan20}
On its own, MP2 can promote fundamental
insights into the physical
properties of chemical systems
that are untarnished by self-interaction errors,
making it a valuable tool
in the arsenal of quantum chemistry.

The \MP\ many-body perturbation series
is based on Rayleigh-Schr\"odinger
perturbation theory (RSPT),\cite{SzaOst82}
which imbues MP2 with the size-consistency and extensivity
that lead to its proper treatment of many-body systems.
On the other hand,
the \MP\ series inherits a divergence problem
from RSPT, such that in the limit of zero-gap systems
the \MP\ correlation energy becomes singular.
While exact degeneracy is perhaps an extreme case
that occurs relatively infrequently in nature,
nonphysically large correlation energies brought on by near-degeneracies
are more commonly encountered.
Large, but not necessarily divergent correlation energies
are often found in systems that exhibit
significant nonadditive correlation effects,\cite{SheRoiLet21}
such as dative bonds in metal complexes\cite{DohHanSte18}
and dispersion-bound complexes
dominated by $\pi$-$\pi$
interactions.\cite{SinValShe02,JafSmi96,TsuUchMat00,S22}
The nonadditive correlation energy can be defined as
the difference between the true correlation energy and
the pairwise correlations captured by MP2,
$E_c^{\text{NA}} = E_c - E_c^{\text{PW}}$.
In cases where MP2 yields poor estimates of the correlation energy,
the nonadditive component is generally large and positive, 
implying that the dominant nonadditive contribution
comes from a screening of the pair correlations.
Indeed, in large systems with extended $\pi$ networks
the MP2 correlation energy becomes catastrophically
large, and without nonadditive screening
interaction energies can be overestimated
by more than 100\%.\cite{CarLaoLiu19,NguCheAge20}

Many useful strategies that account for nonadditive
electron correlation have been developed over the years.
A simple one is to directly scale
the same-spin and\slash or opposite-spin correlation
energies,\cite{Gri03,JunLocDut04,LocShaHea07,LocJunHea05}
which can improve the performance of MP2
for thermochemistry and
noncovalent interactions.\cite{NeeSchKos09}
Another strategy is to use only the short-range
part of the Coulomb operator when evaluating the MP2
energy, thereby attenuating the
range of the correlation interaction
and improving results for a wide range of
chemical problems.\cite{GolDutHea13,GolHea12,GolHea14a,GolBelHea15}
However, while these approaches
treat the symptoms of a completely pairwise
correlation energy approximation,
they do not directly address the underlying cause.

One approach 
that offers direct
screening of pair correlations 
is regularized MP2.
Broadly speaking, regularization modifies the
MP2 energy expression with a function that
damps any divergent or excessively large correlations,
ideally while retaining the unvarnished MP2 energy for weaker
correlations.
Regularization has been used to avoid
singular correlation energies that are encountered while
optimizing molecular orbitals
under a potential that contains the MP2
energy  (orbital-optimized
MP2),\cite{NeeSchKos09,StuHea13,ShaStuSun15,SoyBoz15,RazStuHea17,LeeHea18,BozUnaAla20}
but even without orbital optimization
regularized MP2 can outperform MP2
across myriad chemical problems.\cite{SheRoiLet21}

Regularized MP2 corrects
the divergent nature of the Rayleigh-Schr\"odinger perturbation series
in zero-gap systems.
Singularities manifest in the second-order RSPT energy,
\begin{equation}
	E^{(2)}_{\text{RS}} = \sum\limits_{k\neq0}\frac{\langle\Phi_0|\hat{V}|\Phi_k\rangle
	\langle\Phi_k|\hat{V}|\Phi_0\rangle}{E_0-E_k}
\end{equation}
in cases of degeneracy, {\em i.e.} when $E_k=E_0$.
More appropriate formulations of perturbation theories
have been developed throughout the years in efforts to sidestep
this divergence problem.
These include retaining the excitation degree
(RE) methods,\cite{FinSta93,Fin06,Fin09}
which define the unperturbed Hamiltonian as
one that is block-diagonal in configuration space
and the perturbation as the couplings between
ancillary excitation blocks. The RE approaches offer
substantial improvements over MP2,
with orbital-optimized RE\slash MP2 approaches
often attaining chemical accuracy for
thermochemical properties.\cite{Fin22}
There has also been substantial effort
to improve many-body perturbation theory
with Green's function based
methods.\cite{LanKanZgi16,NeuBaeZgi17,CovTew23}


A less modern approach
that has regained considerable attention in recent
years was pioneered
in the 1930s by Lennard-Jones, Brillouin, and Wigner
as an
alternative to the Rayleigh-Schr\"odinger power series
and came to be known as Brillouin-Wigner (or Lennard-Jones-Brillouin-Wigner)
perturbation theory (BWPT).\cite{Len30,Bri32,Wig35,HubWil10}
The first term where BWPT differs from RSPT is the
second-order energy, which takes the form,
\begin{equation}\label{eq:BWPT}
	E^{(2)}_{\text{BW}} = \sum\limits_{k\neq0}\frac{\langle\Phi_0|\hat{V}|\Phi_k\rangle
	\langle\Phi_k|\hat{V}|\Phi_0\rangle}{E_0-E_k+E^{(2)}_{\text{BW}}}
\end{equation}

There are a few distinct advantages to BWPT:
it converges more rapidly than RSPT for a given problem\cite{Bri32,Wig35}
and it is regular at all orders due to $E^{(n)}_{\text{BW}}$ in the
denominator. In fact, second-order BWPT is exact for a two-level system
while RSPT requires summation to infinite order to achieve the exact
result.\cite{HubWil10}
On the other hand, $E^{(2)}_{\text{BW}}$ appears on both
sides of the above expression and must
therefore be determined self-consistently.
While this does increase the cost of
the perturbation theory, it is not the fatal flaw that has limited
the application of BWPT in quantum chemistry over the last half century.
Instead,
BWPT fell into disuse after
it was found that it 
is not size-extensive and therefore fails as a proper many-body
theory.\cite{MarYouSam67}

Despite its failures for single-reference systems,
the mathematical form of the Brillouin-Wigner series is convenient
for multireference theories and is still actively
used in this
context.\cite{Wen98,MasHubMac98,PitNacCar01,SinChaSud10,ManRayCha19,Cha20,Cha21} 
In particular, it is notable that the
Brillouin-Wigner cluster expansion of the wave function
is equivalent to the Rayleigh-Schr\"odinger one
with the key exception that
multireference Brillouin-Wigner coupled-cluster
theory is immune to the intruder state problem.\cite{HubNeo94}
The treatment of intruder states and the divergences 
encountered in single-reference perturbation theories
are closely linked,\cite{BatFraFde22}
so it is natural to wonder whether the problems
in single-reference BWPT can be amended to obtain
a regular correlation energy at MP2 cost.

If BWPT could be made size-consistent and size-extensive,
it could supply correlation energies that naturally incorporate
nonadditive screening effects at all orders.
This has spurred interest in deriving size-extensivity corrections
for BWPT from the Bloch equations,\cite{Pit03}
and through renormalization of the second-order
energy.\cite{Aks12}
Recently,
an alternative
{\em ansatz} to standard BWPT was proposed,\cite{KelTsaReu22}
where the correlation energy per electron ($E_{\text{BW}}^{(2)}/N_{e}$)
was inserted into the denominator of Eq.~\ref{eq:BWPT}
in an effort to restore size-extensivity.
Importantly, Ref.~\citen{KelTsaReu22} pointed out
that the derivation of Eq.~\ref{eq:BWPT} can be generalized to
an arbitrary level-shift in place
of $E_0+E_{\text{BW}}^{(2)}$,
thus opening the door for a wide variety of level-shift
energies to be conceived and applied.

In this work, we present a different approach to this problem, based on a partitioning of the Hamiltonian
that incorporates a judiciously designed one-electron regularization operator
into the zero-order Hamiltonian while the remainder
of the correlation energy is described as a perturbation.
Furthermore, we cast the second-order energy expression
into a tensor framework,
ensuring that our approach
retains invariance to unitary transformations among
the occupied or virtual orbitals.
Our chosen form of the regularization
operator
satisfies size-consistency and extensivity through second order.
We benchmark the performance
of our proposed method across a wide
variety of datasets
where MP2 performs rather poorly,
including covalent bond breaking, noncovalent interaction
energies, reaction barrier heights, thermochemical properties,
and metal\slash organic reaction energies.


\section{Theory}
Throughout this work, \{$i, j, k$\dots\} refer
to occupied orbitals, \{$a, b, c$\dots\} refer to
unoccupied orbitals, \{$p, q, r$\dots\} are
arbitrary orbitals, and \{$P, Q, R,$\dots\} 
are auxiliary functions.

\subsection{Orbital-Energy Dependent Regularized MP2}
The MP2 correlation energy in the canonical molecular orbital basis is,
\begin{equation}\label{eq:MP2Energy}
	E_c = -\frac{1}{4}\sum\limits_{ijab}\frac{|\mathbb{I}_{ijab}|^2}{\varepsilon_a+\varepsilon_b-\varepsilon_i-\varepsilon_j} = -\frac{1}{4}\sum\limits_{ijab}\frac{|\mathbb{I}_{ijab}|^2}{\Delta_{ij}^{ab}} \; ,
\end{equation}
where,
\begin{equation}
	\mathbb{I}_{ijab} = (ij||ab)
\end{equation}
are the antisymmetrized two-electron integrals and $\varepsilon_p$ is
the orbital energy of orbital $p$.
This expression for the correlation energy is clearly divergent when
the denominator approaches zero, but the energy may
become much too large long before this limit is reached
if nonadditive screening is particularly important.

\begin{table*}[ht!!]
\caption{Various choices for $E_{\text{LS}}$ in Eq.~\ref{eq:GeneralBW} and properties of the resultant correlation energy.}\label{table:Functionals}
\begin{center}
\begin{tabular}{l c c c c}
\hline\hline
\mc{1}{c}{Method} & \mc{1}{c}{$E_{\text{LS}}$} & \mc{1}{c}{Size-consistent}
& \mc{1}{c}{Size-extensive} & \mc{1}{c}{Invariant}\\ 
\hline
MP2, $\kappa$-MP2, $\sigma^p$-MP2 & $E_0$ & \textcolor{dgreen}{\cmark} & \textcolor{dgreen}{\cmark} & \textcolor{dgreen}{\cmark} \\[5pt]
$\delta$-MP2 & $E_0+\delta$; $\delta>0$ & \textcolor{dgreen}{\cmark} & \textcolor{dgreen}{\cmark} & \textcolor{dgreen}{\cmark} \\[5pt]
IEPA\slash BGE2 & $E_0+e_{ij}$; $e_{ij} = -\frac{1}{4}\sum\limits_{ab}\frac{|(ij||ab)|^2}{\Delta_{ij}^{ab} + e_{ij}}$
& \textcolor{dgreen}{\cmark} & \textcolor{Red}{\xmark} & \textcolor{Red}{\xmark}\\[10pt]
BW2 & $E_0+E_c^{\text{BW2}}$ &  \textcolor{Red}{\xmark} &  \textcolor{Red}{\xmark} & \textcolor{dgreen}{\cmark}\\[5pt]
xBW2 & $E_0+E_c^{\text{BW2}}/N_e$ &  \textcolor{Red}{\xmark} & \textcolor{dgreen}{\cmark} & \textcolor{dgreen}{\cmark}\\[5pt]
BW-s2$^a$ & $\bar{E}_0+E^{(2)}-E_{\text{R},k}$
& \textcolor{dgreen}{\cmark} & \textcolor{dgreen}{\cmark} & \textcolor{dgreen}{\cmark}\\
\hline\hline
\mc{5}{l}{\footnotesize
$^a$Second-order size-consistent Brillouin-Wigner perturbation theory with shifted $\hat{H}_0$ (BW-s2); this work.
}
\end{tabular}
\end{center}
\end{table*}

A straightforward approach to tempering this bad behavior
is to add a level-shift to the denominator of the form
$\Delta_{ij}^{ab} + \delta$ where $\delta>0$,\cite{StuHea13,ShaStuSun15,RazStuHea17}
but this approach generally provides too weak of regularization
and lacks input from the underlying physics of the system.
More sophisticated regularizers that have orbital energy dependence
can be derived by Laplace transform of Eq.~\ref{eq:MP2Energy}
where the correlation energy can be exactly rewritten as,\cite{Alm91}
\begin{equation}
	E_c = -\frac{1}{4}\sum\limits_{ijab}\int_0^\infty d\tau e^{-\tau\Delta_{ij}^{ab}}|\mathbb{I}_{ijab}|^2
\end{equation}
From here, the upper integration bound can be truncated to a finite value,
$\sigma(\Delta_{ij}^{ab})^{p-1}$ to give,
\begin{equation}
	E_c = -\frac{1}{4}\sum\limits_{ijab} \frac{|\mathbb{I}_{ijab}|^2}{\Delta_{ij}^{ab}}
	\Big(1 - e^{-\sigma(\Delta_{ij}^{ab})^p}\Big)\; ,
\end{equation}
where $p=1$ gives what is known as $\sigma$-MP2.
The case $p=2$ gives $\sigma^2$-MP2 and can be derived
through second-order perturbative analysis of the
flow equations.\cite{Eva14b,WanLiEva19}

In this work, we will focus on a flavor of empirical regularization
known as $\kappa$-MP2,\cite{LeeHea18}
where the integrals themselves are
damped by a factor of $(1-\text{exp}[-\kappa\Delta_{ij}^{ab}])$ leading to,
\begin{equation}
	E_c = -\frac{1}{4}\sum\limits_{ijab} \frac{|\mathbb{I}_{ijab}|^2}{\Delta_{ij}^{ab}}
	\Big(1 - e^{-\kappa\Delta_{ij}^{ab}}\Big)^2
\end{equation}
All of the above orbital-energy dependent ($\Delta$-dependent) regularizers rely on a single
empirical parameter ($\sigma$ or $\kappa$) that is somewhat transferable,
but expresses different optimal values for different classes of chemical problem.\cite{SheRoiLet21}
We will limit our investigations in this work to $\kappa$-MP2, but
given that all of the aforementioned flavors of $\Delta$-dependent regularization
yield similar results,\cite{SheRoiLet21}
we expect the conclusions drawn here for $\kappa$-MP2
to be general for this class of regularizer.

\subsection{Brillouin-Wigner Theory With Modified Energy}
It was recently proposed that the
second-order Brillouin-Wigner energy
can be derived as a specific case of the
more general correlation expression,\cite{KelTsaReu22}
\begin{equation}\label{eq:GeneralBW}
	E^{(2)} = \sum\limits_{k\neq0}
	\frac{\mybra{\Phi_0}\hat{V}\myket{\Phi_k}\mybra{\Phi_k}\hat{V}\myket{\Phi_0}}{E_{\text{LS}}-E_k}\; ,
\end{equation}
where $E_{\text{LS}}$ is an arbitrary level-shift.
Usually, $E_{\text{LS}}$ is taken to be the exact
ground-state energy, $E_{\text{LS}}=E$,
but the consideration of a more general $E_{\text{LS}}$
unlocks myriad possibilities for the precise form of
the correlation energy.
In effect, this
reframes the BWPT problem in terms of
$E_c[E_{\text{LS}}(\Psi_0)]$,
where the correlation energy is expressed in terms
of a level-shift energy that itself depends on the wave function.
%
Inserting various {\em ans{\"a}tze} into
Eq.~\ref{eq:GeneralBW} leads to different correlation energies.
For example, setting $E_{\text{LS}} = E_0$ yields second-order
M{\o}ller-Plesset perturbation theory and
$E_{\text{LS}}=E_0+\delta$ gives $\delta$-MP2.
Other choices include the pair-correlation
energy ($E_{\text{LS}}=E_0+e_{ij}$),
which leads to the independent electron pair approximation (IEPA)
or the second-order Bethe-Goldstone equation (BGE2),\cite{SzaOst82,ZhaRinSch16,ZhaRinPer16}
the second-order correlation energy ($E_{\text{LS}}=E_0+E^{(2)}$)
gives second-order BWPT, and
the correlation energy per electron ($E_{\text{LS}}=E_0+E^{(2)}/N_e$)
gives the size-extensive xBW2 method.\cite{KelTsaReu22}
Each choice results in a different correlation energy with different
mathematical properties that are summarized
in Tab.~\ref{table:Functionals}.

\subsection{Repartitioned Brillouin-Wigner Perturbation Theory}
Inspired by the generality of such a modification to BWPT,
we consider a slightly more formalized approach by partitioning
the Hamiltonian such that the zero-order Hamiltonian
contains
a regularizing operator that
modulates the occupied orbital energies.
Specifically, we propose the following partition,
\begin{equation}
	\hat{H} = \hat{\bar{H}}_0 + \lambda\hat{\bar{V}} \; ,
\end{equation}
where,
\begin{equation}\label{eq:ShiftedH0}
	\begin{split}
	\hat{\bar{H}}_0 &= \hat{H}_0 + \hat{R}\\
	\hat{\bar{V}} &= \hat{V} - \hat{R}
	\end{split}
\end{equation}
where $\hat{H}_0$ is the Fock operator,
$\hat{\bar{V}}$ contains all of the many-body
correlations that are not contained within $\hat{H}_0$ and $\hat{R}$,
and $\hat{R}$ is a one-electron regularizer operator
of the form,
\begin{equation}
	\hat{R} = \sum\limits_{ij}r_{ij}a_j^\dagger a_i \; .
\end{equation}
Of particular note is the fact that
the infinite summation of
the Brillouin-Wigner perturbation series
is invariant to partitioning
the Hamiltonian in this way.\cite{Fee56}

Next, we write the perturbed Schr{\"o}dinger equation as
\begin{equation}
	(E-\hat{\bar{H}}_0)|\Psi\rangle=\lambda\hat{\bar{V}}|\Psi\rangle
\end{equation}
Defining $\hat{Q}=1-|\Phi_0\rangle\langle\Phi_0|$
and multiplying by this quantity on the left we find,
\begin{equation}
	\hat{Q}|\Psi\rangle = \lambda\hat{Q}(E-\hat{\bar{H}}_0)^{-1}\hat{\bar{V}}|\Psi\rangle
	= \lambda\hat{\Gamma}_0\hat{\bar{V}}|\Psi\rangle
\end{equation}
where,
\begin{equation}
	\hat{\Gamma}_0=\sum\limits_{k\neq0}\frac{|\Phi_k\rangle\langle\Phi_k|}{E-\bar{E}_k} 
\end{equation}
is the resolvent.
In the above,
we have assumed that $\Phi_k$ are also eigenfunctions of $\hat{H}_0$,
such that,
\begin{equation}\label{eq:H0}
	\bar{E}_k=\langle\Phi_k|\hat{\bar{H}}_0|\Phi_k\rangle =
	\sum\limits_i^{\text{occ}} \big(F^i_{\bullet i} + R^i_{\bullet i}\big) 
\end{equation}
where $\bar{E}_k = E_k + E_{\text{R},k}$, is the
energy of state $k$ as modulated by the
regularizer operator.

Taking the usual assumption of
intermediate normalization, {\em i.e.} $\langle\Phi_0|\Psi\rangle=1$,
allows us to
expand the wave function and energy in a perturbation series,
\begin{equation}
\begin{split}
	\Psi^{(n)} &= \sum\limits_{m=0}^n (\lambda\hat{\Gamma}_0\hat{\bar{V}})^m|\Phi_0\rangle\\
	E^{(n)} &= \lambda\langle\Phi_0|\hat{\bar{V}}|\Psi^{(n-1)}\rangle
\end{split}
\end{equation}
Therefore, to first order in $E$, we find,
\begin{equation}
	\begin{split}
	E^{(1)} &= \langle\Phi_0|\hat{\bar{V}}|\Psi^{(n-1)}\rangle\\
	&= \langle\Phi_0|\hat{H}|\Phi_0\rangle -  \langle\Phi_0|\hat{\bar{H}}_0|\Phi_0\rangle
	 = E_{\text{HF}} - \bar{E}_0
	\end{split}
\end{equation}
which when combined with Eq.~\ref{eq:H0} (for $k=0$) gives the
usual result for the first-order energy,
$E = \bar{E}_0+E^{(1)} = E_{\text{HF}}$. Thus, there is
no first-order correction to the Hartree-Fock energy, $E_{\text{HF}}$.

The second-order correction differs from
BWPT and RSPT,
\begin{equation}\label{eq:PT2Energy}
\begin{split}
E^{(2)} &= \sum\limits_{k\neq0}\frac{\langle\Phi_0|\hat{V}|\Phi_k\rangle
	\langle\Phi_k|\hat{V}|\Phi_0\rangle}{E-\bar{E}_k}\\
	&= \sum\limits_{k\neq0}\frac{\langle\Phi_0|\hat{V}|\Phi_k\rangle
	\langle\Phi_k|\hat{V}|\Phi_0\rangle}{(E_0-E_k)+(E_{\text{R},0}-E_{\text{R},k})+E^{(2)}}
\end{split}
\end{equation}
where $E=\bar{E}_0+E^{(2)}$ has been substituted in the denominator,
as per the usual BWPT approach.
The fully expanded denominator
now consists of the zero-order energy gap
$E_0-E_k$, the correlation energy $E^{(2)}$, and a
new contribution from the regularizer $E_{\text{R},0}-E_{\text{R},k}$ that changes the state energies.
Interestingly, if we take $\bar{E}_0+E^{(2)}-E_{\text{R},k}=E_{\text{LS}}$
in accordance with the proposed approach in Ref.~\citen{KelTsaReu22}, we recover
Eq.~\ref{eq:GeneralBW}.

\subsection{Tensor formulation of the second-order energy}
A convenient tool that ensures orbital invariance
of our final correlation energy expression is
the tensor formulation of many-body perturbation
theory.\cite{HeaMasWhi98,LeeMasHea00,DiSJunHea05}
For MP2, the linear
amplitude equation takes the form,
\begin{equation}\label{eq:MP2tensor}
	\sum\limits_{klcd}\Delta_{ijkl}^{abcd}\cdot t_{kl}^{cd} = -\mathbb{I}_{ijab} \; ,
\end{equation}
where 
$t_{kl}^{cd}$ are the amplitudes, and
\begin{equation}\label{eq:DeltaTensor}
	\Delta_{ijkl}^{abcd} = (F_{ac}\delta_{bd} + \delta_{ac}F_{bd})\delta_{ik}\delta_{jl}
	- (F_{ik}\delta_{jl} + \delta_{ik}F_{jl})\delta_{ac}\delta_{bd}
\end{equation}
is the usual 8-rank tensor composed of
Fock matrix elements, $F_{pq}$.
In the basis of canonical molecular orbitals,
where the Fock matrix is diagonal and the orbitals form an orthonormal set,
Eq.~\ref{eq:DeltaTensor} is trivially diagonal such that solving
Eq.~\ref{eq:MP2tensor} leads to
the well-known form of the MP2 amplitudes,
\begin{equation}
	t_{ij}^{ab} = -\frac{\mathbb{I}_{ijab}}{\varepsilon_a+\varepsilon_b-\varepsilon_i-\varepsilon_j}
\end{equation}
which gives way to the usual MP2 energy expression
in Eq.~\ref{eq:MP2Energy}.

Within this framework, the shifted zero-order Hamiltonian
from Eq.~\ref{eq:ShiftedH0} leads to,
\begin{equation}\label{eq:MFCtensor}
\sum\limits_{klcd}\big(\Delta_{ijkl}^{abcd} + R_{ijkl}^{abcd}\big)\cdot t_{kl}^{cd} = -\mathbb{I}_{ijab} 
\end{equation}
where $\mathbf{R}$ is a regularizing tensor.
In the hypothetical
case of diagonal $\bm{\Delta}$ and $\mathbf{R}$
tensors, the amplitudes,
\begin{equation}
t_{ij}^{ab} = -\frac{\mathbb{I}_{ijab}}{\Delta_{ij}^{ab}+R_{ij}^{ab}}
\end{equation}
result in the same energy expression as that of Eq.~\ref{eq:PT2Energy}
with $\mathbf{R}$ playing the role of $E_{\text{R},0}-E_{\text{R},k}$.
To retain size-consistency at second order,
it is crucial
to choose a form of $\mathbf{R}$
that cancels $E^{(2)}$ while still modulating
the orbital energy gap to avoid divergences.

An important feature of Eq.~\ref{eq:PT2Energy}
is that it enables a straightforward mechanism for cancelling out
the redundant correlation terms in the denominator that result in
size-consistency errors in standard BWPT.
Namely, if we can define $\hat{R}$ such that
$\langle\Phi_{ij}^{ab}|(E-\hat{H}_0-\hat{R})|\Phi_{kl}^{cd}\rangle=\Delta_{ijkl}^{abcd}+R_{ijkl}^{abcd}$, then the contributions to the
denominator of the resolvent that arise from the correlation energy
of the entire system
({\em i.e.} $E^{(2)}$ at second order) will vanish,
thereby eliminating size-inconsistent terms.
 To this end, we choose a form of $\mathbf{R}$ that ensures that
the correlation between any two orbitals $\{i, j\}$ goes to
zero when the orbitals are far apart,
\begin{equation}\label{eq:Rtensor}
	R_{ijkl}^{abcd} = \frac{1}{2}(W_{ik}\delta_{jl} + \delta_{ik}W_{jl})\delta_{ac}\delta_{bd} \; ,
\end{equation}
where
\begin{equation}\label{eq:Wmatrix}
	W_{ij} = \frac{1}{2}\sum\limits_{kab} \left[ t_{ik}^{ab}(jk||ab) + t_{jk}^{ab}(ik||ab) \right] \; .
\end{equation}
An important property of $\mathbf{W}$
is that $\text{tr}(\mathbf{W})=E^{(2)}$, which results
in total cancellation of $E^{(2)}$ in the resolvent,
leading to a size-consistent energy expression at second-order.
Specifically, it can be shown that,
\begin{equation}
\langle\Phi_{ij}^{ab}|\hat{R}|\Phi_{kl}^{cd}\rangle=
\sum\limits_n W_{nn} + \frac{1}{2}(W_{ik}\delta_{jl}+\delta_{ik}W_{jl})\delta_{ac}\delta_{bd}
\end{equation}
thus straightforwardly cancelling $E^{(2)}$ while modifying the matrix elements connecting pairs of occupied orbitals (and the occupied-orbital energies).
However, we note that size-inconsistent terms enter at
third and higher orders. 


Matrix elements of Eq.~\ref{eq:Wmatrix} appear also in an orbital invariant
CEPA(3) correction,\cite{NooLeR06}
and share similarities with one of the terms in the
MP2 orbital energy gradient.\cite{LeeHea18}
In particular, Eq.~\ref{eq:Wmatrix} is related to
the correlation contribution to the ionization energy of orbital $i$,
\begin{equation}\label{eq:IPcorr}
	E_c^{\text{IP,}i} = \frac{1}{2}\sum\limits_{kab}t_{ik}^{ab}(ik||ab)
\end{equation}
where the orbitals are fixed at those of the $n$-electron system.
One may notice that the elements of $\mathbf{W}$
in Eq.~\ref{eq:Wmatrix} correspond to $2E_i^{\text{IP,corr}}$.
Not only does this factor of 2 naturally emerge from
the necessity of cancelling $E^{(2)}$ in the resolvent, but
it can also be understood as a means of modulating
the energies of both occupied orbitals involved in any double substitution
with $E_c^{\text{IP},i}$. We elaborate further on this point
in Appendix~A.

One complication that arises in the solution of Eq.~\ref{eq:MFCtensor}
with our proposed form of $\mathbf{R}$
is that in the canonical orbital
basis the left-hand side of Eq.~\ref{eq:MFCtensor} is not diagonal.
Instead, it takes the form,
\begin{equation}
\begin{split}
	&\sum_{klcd}\bigg\{[F_{ac}\delta_{bd}+\delta_{ac}F_{bd}]\delta_{ik}\delta_{jl}
	-\delta_{ac}\delta_{cd}[F_{ik}\delta_{jl}+\delta_{ik}F_{jl}]\\
	&\qquad - \frac{\delta_{ac}\delta_{bd}}{2}(W_{ik}\delta_{jl} + \delta_{ik}W_{jl})\bigg\}t_{kl}^{cd}
	= -\mathbb{I}_{ijab}
\end{split}
\end{equation}
which, after contracting
the first two terms over all orbital indexes $\{k, l, c, d\}$
and the final four terms
over virtual-orbital indexes $\{c, d\}$
gives,
\begin{equation}\label{eq:MFCsimple}
\begin{split}
	&[\varepsilon_{a}+\varepsilon_{b}]t_{ij}^{ab}\\
	&\qquad-\sum_{kl}\bigg[\bigg(F_{ik}+\frac{W_{ik}}{2}\bigg)\delta_{jl}+\delta_{ik}\bigg(F_{jl}+ \frac{W_{jl}}{2}\bigg)\bigg]t_{kl}^{ab}\\
	&\qquad\qquad= -\mathbb{I}_{ijab}
\end{split}
\end{equation}
Where we have not carried out the contraction over
indexes $k$ and $l$ for the occupied-occupied
block of $(\bm{\Delta}+\mathbf{R})\cdot\mathbf{t}$
as to emphasize both that $\mathbf{W}$
only changes the occupied-occupied
block and that $\mathbf{W}$ is not diagonal
in the basis of canonical orbitals.

One way to solve Eq.~\ref{eq:MFCsimple}
is to store the amplitudes in memory and
solve for them using an iterative scheme,
as is often done in local correlation methods.\cite{MasHea98,Pul83,SaePul93}
However, amplitude storage can be avoided if we
find a suitable basis wherein the left-hand side of
Eq.~\ref{eq:MFCsimple} is diagonal.
To accomplish this goal, we can leverage
the orbital invariance of Eq.~\ref{eq:MFCsimple}
by rotating the occupied molecular orbitals
into a basis where the matrix $\mathbf{F}_{\text{oo}}+\frac{1}{2}\mathbf{W}$
is diagonal (where $\mathbf{F}_{\text{oo}}$ is the occipied-occupied
block of the Fock matrix).
To find the appropriate rotation, we solve the
Hermitian eigenvalue equation,
\begin{equation}
	\bigg(\mathbf{F}_{\text{oo}}+\frac{1}{2}\mathbf{W}\bigg)\mathbf{U}
	= \tilde{\varepsilon}\mathbf{U}
\end{equation}
where $\tilde{\varepsilon}$ are a set of {\em dressed}
occupied orbital eigenvalues.
Rotating the occupied molecular orbital coefficients, $\mathbf{C}_{\text{occ}}$,
into this new basis {\em via} the unitary matrix, $\mathbf{U}$,
\begin{equation}\label{eq:BasisTransform}
	\mathbf{\tilde{C}}_{\text{occ}} = \mathbf{C}_{\text{occ}}\mathbf{U}
\end{equation}
ensures that the tensor $\bm{\Delta}+\mathbf{R}$ is diagonal.
In this new basis, Eq.~\ref{eq:MFCsimple} takes the form,
\begin{equation}\label{eq:DressedBasisAmplitudes}
	(\varepsilon_a + \varepsilon_b - \tilde{\varepsilon}_i - \tilde{\varepsilon}_j)\tilde{t}_{ij}^{ab}
	= -\tilde{\mathbb{I}}_{ijab}
\end{equation}
where the integrals in $\tilde{\mathbb{I}}_{ijab}$ have been rotated into the
new basis.
Solving the transformed equation gives the amplitudes,
\begin{equation}
	\tilde{t}_{ij}^{ab}
	= -\frac{\tilde{\mathbb{I}}_{ijab}}{(\varepsilon_a + \varepsilon_b - \tilde{\varepsilon}_i - \tilde{\varepsilon}_j)}
\end{equation}
and the energy
\begin{equation}\label{eq:MP2like}
	\tilde{E}_c
	= -\frac{1}{4}\sum\limits_{ijab}\frac{|\tilde{\mathbb{I}}_{ijab}|^2}{(\varepsilon_a + \varepsilon_b - \tilde{\varepsilon}_i - \tilde{\varepsilon}_j)}
\end{equation}

Note the use of the dressed eigenvalues $\tilde{\varepsilon}_p$
in the above equations, which are a consequence of the change of basis.
These dressed eigenvalues are modulated by the choice of $\mathbf{R}$,
which in our case is related to the ionization potential of the orbital.
Specifically, using Koopmans' theorem\cite{Koopmans34}
we may rewrite the canonical orbital-energy differences as,
\begin{equation}
	\Delta_{ij}^{ab} = E^{\text{IP}}_i + E^{\text{IP}}_j - E^{\text{EA}}_a - E^{\text{EA}}_b
\end{equation}
where $E^{\text{IP}}_p$ and $E^{\text{EA}}_p$ are the ionization energy
and electron affinity of orbital $p$, respectively.
Considering the relationship in Eq.~\ref{eq:IPcorr},
the action of our regularizer is to
replace $E^{\text{IP}}_p$ with their correlated counterparts,
$\tilde{E}^{\text{IP}}_i=E^{\text{IP}}_i + E^{\text{IP,corr}}_i$,
thus augmenting the gap by correlating the ionization energies.
Notably, this concept of correcting the
quasiparticle energies has strong similarities to Green's function
based perturbation theories\cite{LanKanZgi16,NeuBaeZgi17}
which are actively being
explored in the context of regularized perturbation
theories.\cite{CovTew23}

Our adherence to the tensorial formalism
and careful consideration of exact conditions
ensures that this {\em ansatz} for the 
form of $\hat{R}$ retains
crucial properties such as size-consistency, size-extensivity,
and orbital invariance in the second order energy.
Therefore, we limit our studies in this work
to those that probe the properties of Brillouin-Wigner perturbation theory with a size-consistent second-order
correlation energy, herein denoted BW-s2.
The size-consistency of BW-s2
can indeed be proven, and we have done so in
Appendix~B.

\begin{figure}
	\centering
	\fig{1.0}{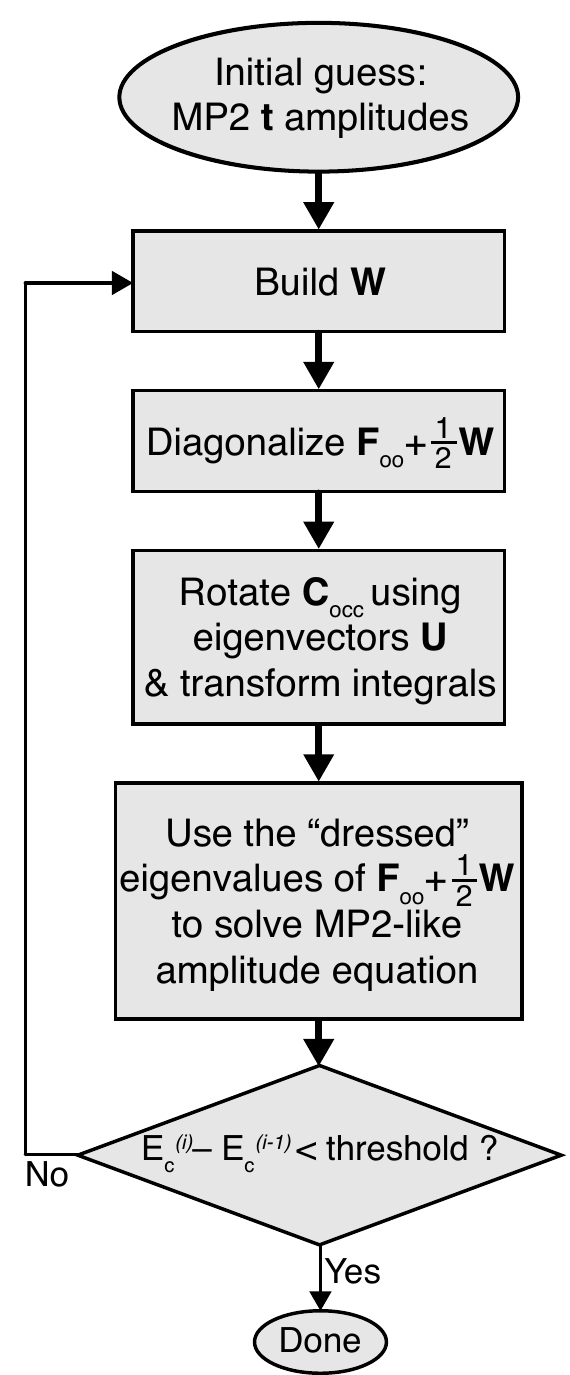}
	\caption{
	Flowchart outlining the iterative procedure
	for solving for the amplitudes for any
	orbital-invariant
	second-order correlation method.
	$E_c^{(i)}$ indicates the correlation energy
	on the current iteration $i$, and
	$E_c^{(i-1)}$ is the correlation energy from
	the previous iteration.
	}\label{fig:Flowchart}
\end{figure}

While we avoid amplitude storage,
the BW-s2 energy expression remains self-consistent
because the $\mathbf{W}$ matrix depends
on the amplitudes, which themselves
depend on the modulation of the energy gap supplied by
the $\mathbf{W}$ matrix.
The flowchart in Fig.~\ref{fig:Flowchart} shows the
iterative protocol that we use to solve for the amplitudes.
We opt for an energy convergence threshold such that
once the change in energy between iterations is sufficiently
small, the algorithm converges.
We note that this procedure is general and can be
used in conjunction with all of the
orbital-invariant methods listed
in Tab.~\ref{table:Functionals}.
As an example, for MP2 the $\mathbf{R}$ tensor is
simply the zero matrix
so the rotations supplied by $\mathbf{U}$
are the identity matrix, $\mathbf{I}$, and the algorithm converges
in one step.
This corresponds
to setting $E=E_0$ in Eq.~\ref{eq:PT2Energy}
with the matrix representation of
$\hat{R}$ being the zero matrix.
Similarly, for $\delta$-MP2, $\mathbf{R}$ is a diagonal matrix
whose nonzero entries are the value of $\delta$, leading
again to a one-step solution.
In the case of the BW2 and xBW2 methods,
the $\mathbf{W}$ matrix is diagonal
with elements $E_c^{\text{BW2}}\delta_{ij}$
or $(E_c^{\text{xBW2}}/N_e)\delta_{ij}$, again leading
to $\mathbf{U}=\mathbf{I}$, but with a self-consistent
energy expression that will still require several cycles to converge.

In order to greatly speed up the evaluation of the two-electron
integrals we use the resolution-of-the-identity (RI) approximation,\cite{FeyFitKom93,BerHar96}
where,
\begin{equation}
	(ia|jb) = \sum\limits_{PQ}(ia|P)(P|Q)^{-1}(Q|jb)
\end{equation}
The RI fit coefficients, $C_{pq}^P$, for the $|pq\rangle$ charge
distribution are,
\begin{equation}
	C_{pq}^P = \sum\limits_{pqQ} (P|Q)^{-1}(Q|pq)
\end{equation}
the 3-center, 2-particle density matrix is,
\begin{equation}
	\Gamma_{ai}^P = \sum\limits_{jb} t_{ij}^{ab}C_{jb}^P
\end{equation}
and finally, we also define
\begin{equation}
	V_{ia}^P = (ia|P) \; .
\end{equation}
This allows us to rewrite Eq.~\ref{eq:Wmatrix} as,
\begin{equation}
	W_{ij} = \frac{1}{2}\sum\limits_{aP} V_{ia}^P\Gamma_{aj}^P
	+ \Gamma_{ia}^PV_{aj}^P
\end{equation}
which is bottlenecked by the ${\mathcal O}(N^5)$ construction of
$\bm{\Gamma}$, therefore adding only trivial overhead to
the usual MP2 energy evaluation.
Finally, we rewrite the MP2-like energy expression from Eq.~\ref{eq:MP2like}
in the dressed-orbital basis as,
\begin{equation}
	\tilde{E}_c = -\frac{1}{2}\sum\limits_{iaP} \tilde{V}_{ia}^P\tilde{\Gamma}_{ai}^P\; ,
\end{equation}
where $\bm{\tilde{\Gamma}}$ and $\bm{\tilde{V}}$
are constructed using the transformed integrals and amplitudes
according to Eqns.~\ref{eq:BasisTransform} and~\ref{eq:DressedBasisAmplitudes}.
With iterative ${\mathcal O}(N^5)$ cost,
the RI approximation makes the BW-s2 approach comparable
in cost to other common methods like CC2,\cite{CC2,HatWei00}
or EOM-MBPT2.\cite{ParPerBar18}

\section{Computational Details}
All calculations were performed in a development version
of Q-Chem~v6.0.2.\cite{QCHEM5}
All SCF convergence thresholds were set to $10^{-8}$
root-mean-square error and the convergence
threshold for the correlation energy
was likewise set to $10^{-8}$~Ha for all calculations except for those
of the L7 dataset, where it
was reduced to $10^{-5}$~Ha for the sake of
computational cost. This should not influence the accuracy of
the calculations because an energy difference of
$10^{-5}$~Ha is only 0.006~kcal\slash mol.
We note that even in the case of a tight Brillouin-Wigner correlation
energy threshold of $10^{-8}$~Ha, only 6 cycles were required
on average (regardless of dataset) to converge the correlation
energy.

To avoid the well-known degradation of perturbation theory results
in systems with appreciable
spin-contamination,\cite{HubCar80,MurDav91,LauStaGau91,AmoAndNic91,KnoAndAmo91,LeeJay93}
we use restricted open-shell
orbitals which are separately pseudocanonicalized in the $\alpha$ and $\beta$ spaces before computing the correlation
energy in all open-shell systems, akin to the RMP2 method.\cite{KnoAndAmo91}
We include the non-Brillouin singles (NBS) contributions
via,
\begin{equation}
	E_{\text{NBS}} = -\sum\limits_{ia}\frac{|F_{ia}|^2}{\varepsilon_a-\varepsilon_i}
\end{equation}
where $F_{ia}$ are off-diagonal Fock matrix elements.
Notably, $E_{\text{NBS}}$ is invariant to the change of basis that is
used to solve the BW-s2 equations.
While Kohn-Sham orbitals have been
applied to M{\o}ller-Plesset perturbation theory with
great effect,\cite{BerLeeHea19,RetHaiBer20,LoiBerLee21}
we emphasize that we use the Hartree-Fock reference determinant throughout this work,
leaving the prospect of combining Kohn-Sham orbitals with BW-s2
for future exploration.

\section{Results and Discussion}
We first assess the fundamental properties of various
second-order correlation methods with some simple numerical tests.
%
We have proven the size-consistency, and by extension
size-extensivity of BW-s2 in Appendix~B, but from a practical
perspective it is important to recover these properties
in numerical calculations, so these tests will
serve to aid in the verification of our implementation.
While we make some conclusions about the size-extensivity, size-consistency,
and unitary invariance of other methods in this section, we emphasize
that these tests are insufficient to prove that a given method has
these properties in general. However, it is necessary that any method
that is size-consistent, size-extensive, and unitary invariant
must recover the expected results
in the following tests, so a failure on any one of these metrics is sufficient to
discount a method from having the property that the metric was designed to test.

For the first test, we check for orbital invariance
by using canonical and Edmiston-Reudenberg localized
orbitals\cite{EdmReu63, EdmReu65} with the cc-pVDZ\cite{Dun89,WooDun94}
basis set on the H$_2$ dimer, placed
in a parallel configuration at 5.4~\AA\ separation.
A method that yields the same correlation energy
despite arbitrary orbital rotations
in the occupied (or virtual) subspace is considered to be orbital invariant,
therefore we should expect no change in the correlation energy
on the change from canonical to
localized orbital representations.
In the cases of the MP2, BW2, xBW2, and BW-s2
methods, the correlation energy remains exactly the
same regardless of orbital representation,
but the IEPA\slash BGE2 method is not invariant
to orbital rotations,
leading to an energy difference of 6.3~meV
between canonical and localized representations.
This is a well known result,\cite{SzaOst82,ZhaRinSch16,ZhaRinPer16}
and actually requires that we skip the orbital rotation step
in the algorithm in Fig.~\ref{fig:Flowchart} for the
IEPA\slash BGE2 method, instead opting for direct solution of the
correlation energy expression with
off-diagonal contributions from the pair-correlation energy in
the denominator.

We next assess size-consistency by calculating the
interaction energy between He and Xe at 40~\AA\ separation
using the Def2-SVP\slash Def2-ECP basis set and
effective core potential.\cite{def2}
A method is considered to be size-consistent
if the total energy for a supersystem comprised of noninteracting subsystems
$A$ and $B$
is the same as the sum of the energies of the individual subsystems,
$E(A+B)=E(A)+E(B)$.
In our case, the He$\cdots$Xe
interaction energy at 40~\AA\ separation should obviously be zero if a
method is properly size-consistent, which is
precisely the result obtained with the MP2, IEPA\slash BGE2,
and BW-s2 method.
The BW2 method has a large residual correlation energy of
111~meV and xBW2 gives a smaller, but still significant
1~meV interaction energy for this system.
This implies that the size-extensive xBW2 method is
not size-consistent, which could have dramatic consequences
in calculations on extended systems, for which it was proposed.\cite{KelTsaReu22}

\begin{figure}[h!!]
\centering
\fig{1.0}{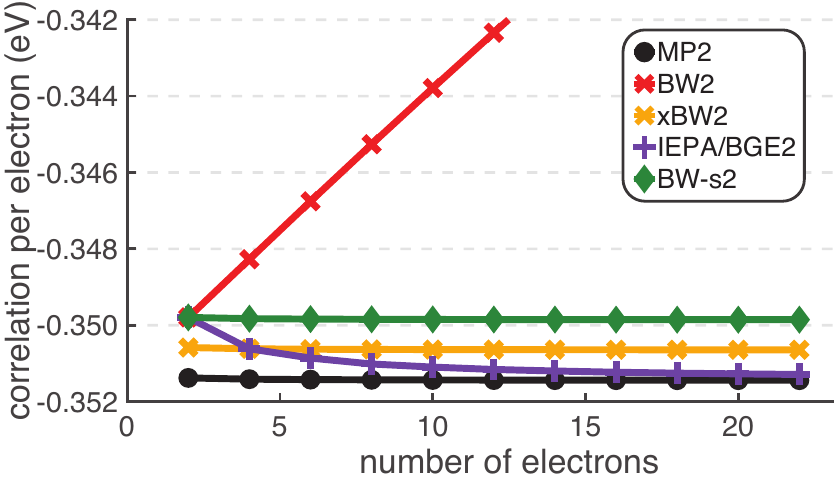}
\caption{
	Correlation energy per electron in He chains
	of increasing size using the cc-pVDZ basis set.
	The spacing between He atoms was set to 3~\AA.
	All calculations were done in the full basis of all
	He atoms by including ghost functions for the atoms
	that are not included explicitly.
}
\label{fig:HeChains}
\end{figure}

Finally, as a metric for size-extensivity Fig.~\ref{fig:HeChains}
examines the correlation energy per electron in a linear chain of He atoms.
A method is considered to be size-extensive
if, for any chain of identical subsystems, the total correlation
energy grows linearly with the number of electrons in the system.
Thus,
the slope of each line in Fig.~\ref{fig:HeChains}
should be zero for a size-extensive method.
This is the case for MP2, xBW2, and BW-s2, but
not for BW2 or IEPA\slash BGE2.
\begin{figure*}[ht!!]
\centering
\fig{1.0}{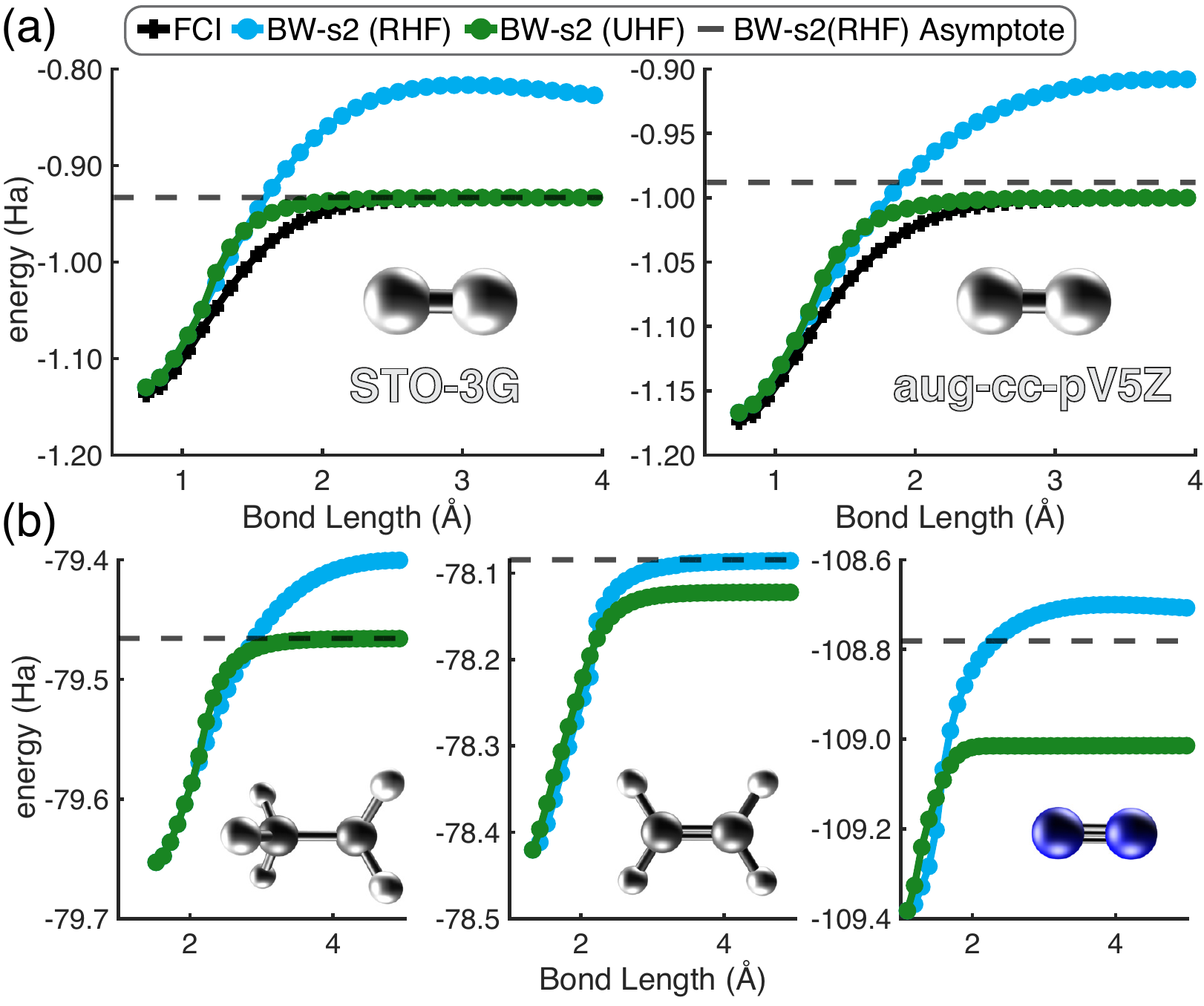}
\caption{
Bond-stretching potential energy curves for (a)
dissociation of the hydrogen molecule
with BW-s2($\alpha=1$) in
(left) a minimal basis set of two orbitals
and (right) a fairly complete basis set, and
(b) from left to right, C--C, C$=$C, and N$\equiv$N
dissociation curves of ethane, ethene, and nitrogen, respectively.
The gray dashed lines mark the asymptotic limit
of BW-s2 with RHF orbitals as numerically estimated
by a calculation in which the bond length was set to 100,000~\AA.
The potential energy curves in (b) were calculated using
the aug-cc-pVQZ basis set.
All equilibrium geometries were optimized at the
$\omega$B97X-V\slash Def2-TZVPPD level\cite{wB97X-V}
and the equilibrium bond distance was incremented by 0.1~\AA\
steps to generate the potential surfaces.
}\label{fig:Dissociation}
\end{figure*}
The Brillouin-Wigner series is infamous for its failure
as a many-body theory, and the monotonic decrease in correlation
energy per electron of BW2 makes this abundantly clear.
IEPA\slash BGE2 exhibits strange behavior,
approaching a slope of zero in the limit of an infinitely large
He chain, but with an inverse-power dependence on the number of electrons.
In this sense, for finite systems IEPA\slash BGE2 is
not size-extensive, and because finite systems encompass
nearly all practical calculations we consider this a
notable failure of
IEPA\slash BGE2.
Lastly, we note that for a single He atom the BW2, IEPA\slash BGE2,
and BW-s2 methods yield the same correlation energy,
which is expected for two-electron systems. 
The summary of the findings of all of these tests can be
seen in Table~\ref{table:Functionals}.

So far, we have used the raw form of Eq.~\ref{eq:Rtensor},
but we should note that this form is somewhat arbitrary
and may be amenable to a scaling parameter, $\alpha$,
that modulates the extent of regularization
in the form $\alpha\mathbf{R}$.
Such a parameter would maintain
the size-consistent/extensive nature of the perturbation
theory, as $E=\bar{E}_0+\alpha E^{(2)}$ in Eq.~\ref{eq:PT2Energy}
and $\text{tr}(\mathbf{W})=\alpha E^{(2)}$ in this case.
Fortunately, the agreement between BW-s2 and BW2 for two-electron
systems offers an exact condition for which
the parameter $\alpha$ can be set.
For a two-state system BWPT yields the exact energy
at second order,\cite{HubWil10}
so we should expect BW2 and a properly parameterized
BW-s2 to achieve the exact dissociation limit for
hydrogen molecule in a minimal (two-orbital) basis set.
It can be shown that in such a minimal basis set
the regularizer tensor in Eq.~\ref{eq:Rtensor}
reduces to the BW2 correlation energy if and only
if $\alpha=1$ ({\em i.e.} the unmodified tensor),
allowing us to set the value of $\alpha$ from first principles.
Somewhat more laboriously, it can also be shown that at the
dissociation limit in
this minimal basis the optimal BW2 and BW-s2 amplitude
is exactly $t_{ii}^{aa}=1$, as expected for a two-electron,
two-orbital system where the orbitals $i$ and $a$ are exactly degenerate.

From here, we resort to numerical testing to illustrate the behavior
of BW-s2($\alpha=1$) for H$_2$ dissociation in minimal and non-minimal
basis sets.
Dissociation curves in both STO-3G\cite{HehStePop69} and
aug-cc-pV5Z basis sets for the $\alpha=1$ case
are shown in Fig.~\ref{fig:Dissociation}a.
As one might expect, the STO-3G results
with restricted Hartree-Fock (RHF) orbitals
show a steep rise to energies that are too high,
while the energies using unrestricted Hartree-Fock (UHF) orbitals
quickly meet the full configuration interaction (FCI)
result for the dissociation limit, leading to the appearance
of a Coulson-Fischer point at 1.3~\AA.
What is most interesting is the behavior at the asymptotic limit,
where the highest occupied molecular orbital
and lowest unoccupied molecular orbital are exactly degenerate
and the RHF MP2 energy diverges.
In this limit, we find that the appropriately parameterized
BW-s2($\alpha=1$)\slash STO-3G
theory converges to the exact FCI result regardless of
whether or not the initial orbitals were spin-polarized.
However, with BW-s2($\alpha=1$, RHF)\slash aug-cc-pV5Z
the FCI limit is no longer attained at second order
and we instead find an upper bound to the exact result.
Encouragingly, the RHF\slash UHF difference remains quite small even in
the large basis set at roughly 12~mHa, so we shall retain the
{\em ab initio} $\alpha=1$
parameter throughout this work.

Repeating this exercise by fitting the value of
$\kappa$ such that the $\kappa$-MP2\slash STO-3G
energy with RHF orbitals equates to the FCI energy at the
asymptotic limit
results in an optimal $\kappa=495.2$~Ha$^{-1}$,
amounting to what appears to be almost no regularization.
However, with a gap of almost zero the $\Delta$-dependent regularizers
naturally suppress most of the correlation energy as the
extent of regularization is proportional to $(1-\text{exp}[-\kappa\Delta_{ij}^{ab}])$,
so the optimal $\kappa$ value must be large to retain any appreciable
amount of correlation.
Such a limit of near-degenerate orbitals therefore
leads to a situation where the optimal value of $\kappa$ becomes
exponentially sensitive to the particular value of the (very small)
energy gap, introducing acute basis set dependence.
For instance, using the same value of $\kappa$
from the STO-3G calculation leads to an RHF-like energy
of $-0.76$~Ha in the aug-cc-pV5Z basis set.
Unfortunately, this implies that such {\em ab initio} parameterization
for $\Delta$-dependent regularizers is not appropriate, but
these results also showcase a crucial advantage of
amplitude-dependent regularization in BW-s2;
namely,
BW-s2 will predict nonzero correlation
energies between orbitals that are exactly
degenerate, perhaps improving its performance
for statically correlated systems.

\begin{figure}[t!!]
\centering
\fig{1.0}{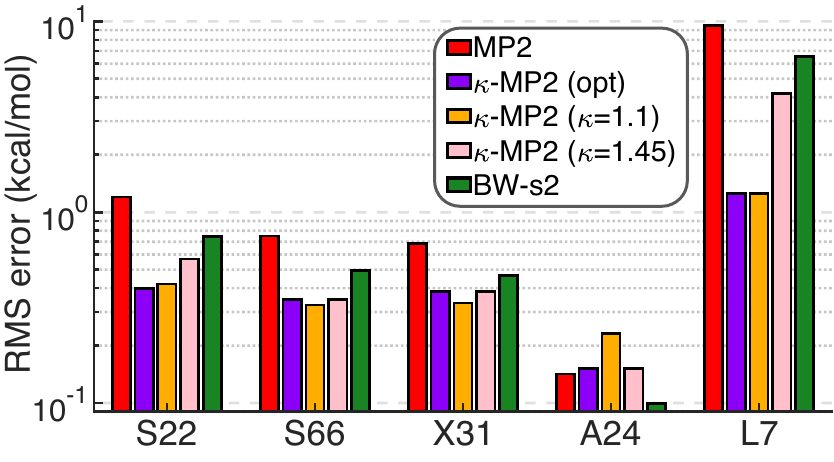}
\caption{
Root-mean-squared errors (log scale) of various second-order
perturbation theories against CCSD(T)\slash CBS
reference energies for noncovalent interaction datasets.
The optimal (opt) $\kappa$ parameters for $\kappa$-MP2
were set to 1.2~Ha$^{-1}$ for S22,
1.45~Ha$^{-1}$ for S66, X31, and A24,
and 1.1~Ha$^{-1}$ for L7 as per Ref.~\citen{SheRoiLet21}.
}
\label{fig:NCIs}
\end{figure}

Additional potential energy curves that feature single-bond breaking
in ethane, double-bond breaking in ethene, and triple-bond breaking
in nitrogen are shown in Fig.~\ref{fig:Dissociation}b.
Remarkably, BW-s2 succeeds in breaking the C\nbd--C sigma bond
in ethane without error in the dissociation limit such that
the RHF and UHF solutions are asymptotically equivalent.
While it might be expected that BW-s2 performs well in the
case of 2-electron, 2-orbital strong correlation, one might be less optimistic
about how a double-substitution theory will hold up when multiple bonds
are dissociated. Indeed, as the bond order increases, the
asymptotic solution of BW-s2 with RHF orbitals deviates further and further from
the spin-polarized result, leading to errors of
47~mHa and 233~mHa for ethene and nitrogen, respectively.
Despite this, the potential energy curves are smooth and don't yield
any particularly surprising results.
The performance of BW-s2 in the sigma-bond breaking
of H$_2$ and ethane
is highly encouraging for future single-bond breaking
applications.


We now turn our attention to the statistical performance
of the BW-s2 method across several noncovalent interaction (NCI) datasets.
The NCI datasets span a wide range of molecular sizes
and interaction types. Where A24,\cite{A24}
S22,\cite{S22} S66,\cite{S66} and the non-I containing subset of
X40 (herein called X31)\cite{X40}
are datasets of small to medium sized nonbonded molecular complexes
with a variety of interaction motifs,
the L7 dataset contains mostly nanoscale $\pi$-stacking interactions
which are particularly difficult for MP2.\cite{L7}
To compare with the benchmark complete basis set limit
(CBS)
coupled-cluster with single, double and perturbative triple
substitutions (CCSD(T)) data,
all perturbation theory results are extrapolated to the CBS
limit using the aug-cc-pVDZ\slash aug-cc-pVTZ\cite{Dun89,WooDun93}
extrapolation scheme from Ref.~\citen{NeeVal11}.
We note for the L7 set that we compare
to the updated domain-localized pair natural orbital
CCSD(T$_0$)\slash CBS\cite{RipNee13,RipSanHan13,RipPinBec16,GuoBecNee18}
benchmarks of Lao and coworkers
and that we use the heavy-aug-cc-pVDZ\slash heavy-aug-cc-pVTZ
extrapolation method that was recommended therein.\cite{VilBalWan22}

\begin{figure}
\centering
\fig{1.0}{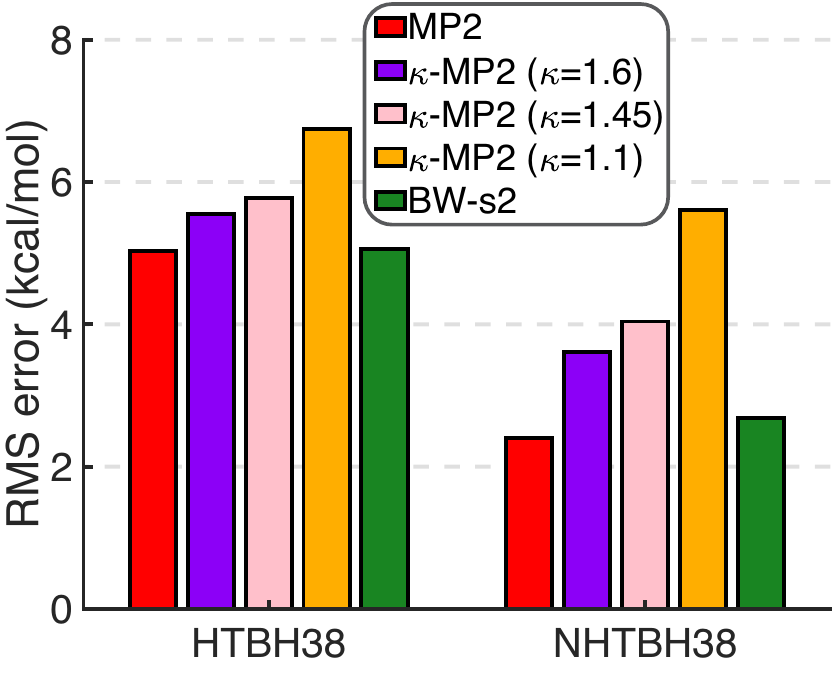}
\caption{
	Root-mean-squared errors of various second-order
	perturbation theories against theoretical best estimate values
	for H-atom transfer (HTBH38) and non-H-atom transfer (NHTBH38)
	datasets.
	All data were extrapolated to the CBS limit using an
	aug-cc-pVTZ\slash aug-cc-pVQZ extrapolation scheme.
}
\label{fig:BHs}
\end{figure}

The results for the NCI datasets in Fig.~\ref{fig:NCIs}
compare BW-s2, MP2,
and $\kappa$-MP2 using several $\kappa$ parameters
for each respective dataset.
Regularized perturbation theories
outperform MP2 for S22, S66,
and X31 datasets, and are only marginally different
from MP2 for the A24 set where MP2 already performs quite well.
The optimal value of $\kappa$ changes a fair amount between
the NCI datasets (1.1$\le\kappa\le$1.45~Ha$^{-1}$),\cite{SheRoiLet21}
so we report results from $\kappa$-MP2 with the optimal
parameter along with the two previously suggested ``universal'' parameters
$\kappa=1.1$\cite{SheRoiLet21} and $\kappa=1.45$.\cite{LeeHea18}
The results obtained with $\kappa=1.1$ and the optimized value of $\kappa$,
$\kappa\text{(opt)}$,
offer consistently low errors across the NCI benchmarks,
and $\kappa=1.45$ performs well across all but the L7 dataset, where the
error increases dramatically
from 1.3 to 4.2~kcal\slash mol with $\kappa\text{(opt)}$
and $\kappa=1.45$, respectively.
Across all NCI datasets, BW-s2 performs
roughly the same as $\kappa=1.45$ with slightly larger
errors on average.
Notably, on the A24 dataset, BW-s2 outperforms all
methods, which contrasts with the fact that all $\kappa$-MP2 results
give errors greater than or equal to MP2. This
suggests that BW-s2 has some degree of flexibility in
its regularization that is not present in $\kappa$-MP2,
perhaps hinting at some additional transferability offered by
the BW-s2 framework.
Overall,
it is encouraging to see such similar performance between BW-s2
and one of the suggested ``universal'' $\kappa$ parameters,
especially given that BW-s2 is parameter-free.

This notion of transferability
can be further tested by examining H and heavy-atom
transfer barrier heights of
HTBH38 and NHTBH38,\cite{ZhaGonTru05,ZheZhaTru07}
where MP2 performs better without regularization.\cite{SheRoiLet21}
The data in Fig.~\ref{fig:BHs}
show that MP2 still performs better without regularization,
but BW-s2 comes very close to this unregularized limit.
As the $\kappa$ parameter in $\kappa$-MP2 is adjusted
away from the rather large (optimal) value of $\kappa=1.6$
to either of the two ``universal'' values of $\kappa=1.45$
or $\kappa=1.1$, the errors climb dramatically.
This is a clear reminder that $\kappa$-MP2 does not
truly have a universal parameter that works well for
all chemical problems, but the good performance of BW-s2
here seems to indicate that BWPT may be more versatile.
The self-consistent nature of the BWPT equations
leads to a modulation of the fundamental gap that
is informed by the value of $\mathbf{W}$,
which in turn is informed by the amplitudes,
introducing a feedback loop that leads to
improved transferability of BW-s2
over that of noniterative
gap-dependent regularizers.

\begin{figure}[t!!]
\centering
\fig{1.0}{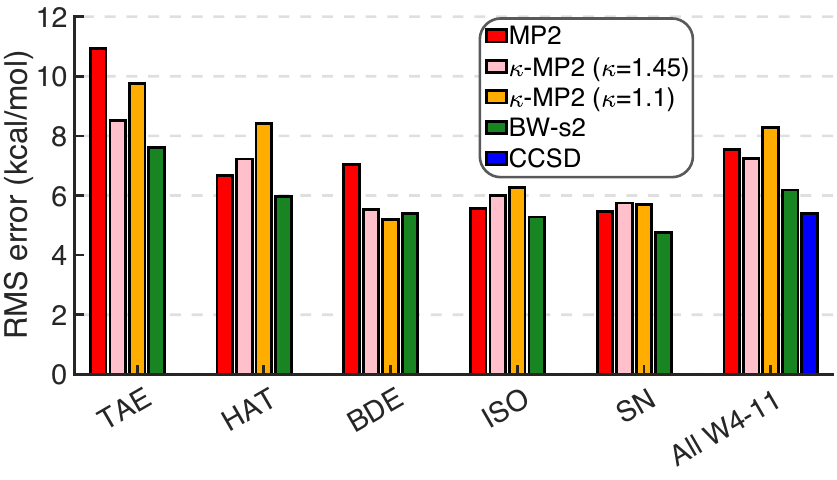}
\caption{
	Root-mean-squared errors for various second-order
	perturbation theories with respect to the subsets
	of thermochemical data within W4-11 which includes
	total atomization energies (TAE), heavy-atom transfer (HAT),
	bond-dissociation energies (BDE), isomerization (ISO), and
	nucleophilic substitution (SN) energies.
	The CBS limit results for all of W4-11
	were obtained with an aug-cc-pVTZ\slash aug-cc-pVQZ
	extrapolation scheme and are compared
	with CCSD data from Ref.~\citen{LeePhaRei22}.
}\label{fig:Thermochem}
\end{figure}

We now consider the nonmultireference subset of
the W4-11 thermochemical database,\cite{W4-11}
which includes 124 atomization energies,
505 heavy-atom transfer energies,
83 bond-dissociation energies,
20 isomerization energies,
and 13 nucleophilic substitution energies
of small molecules and radicals.
The data in Fig.~\ref{fig:Thermochem}
show that $\kappa$-MP2 does not generally improve upon
the MP2 results, occasionally making matters worse
for heavy-atom transfers, isomerization, and nucleophilic substitution
energies. Even the value of $\kappa=1.45$, which was
parameterized on the W4-11 dataset, performs only about as well
as MP2 overall.
On the other hand, BW-s2 far exceeds the performance of
$\kappa$-MP2
with tangible reductions in errors across all subsets of W4-11
except for bond-dissociation energies which remain roughly the same.
For the whole W4-11 set,
the BW-s2 results are markedly better than MP2
and $\kappa$-MP2,
even rivaling the overall performance of CCSD.

The largest improvements offered by BW-s2 are in
the total atomization energies, improving upon the
MP2 results by 3~kcal\slash mol.
Overall, BW-s2 has a root-mean-squared error (RMSE)
of 6.2~kcal\slash mol for W4-11, improving thermochemical properties
relative to MP2 and $\kappa$-MP2 by roughly 1.5~kcal\slash mol.
These data suggest that the BW-s2 method does not
only track well with gap-dependent regularizers
for NCIs,
but for barrier heights and general thermochemical properties
it exceeds their performance,
implying that BWPT approaches may be more
transferable across chemical problems.

\begin{figure}[t!!]
\centering
\fig{1.0}{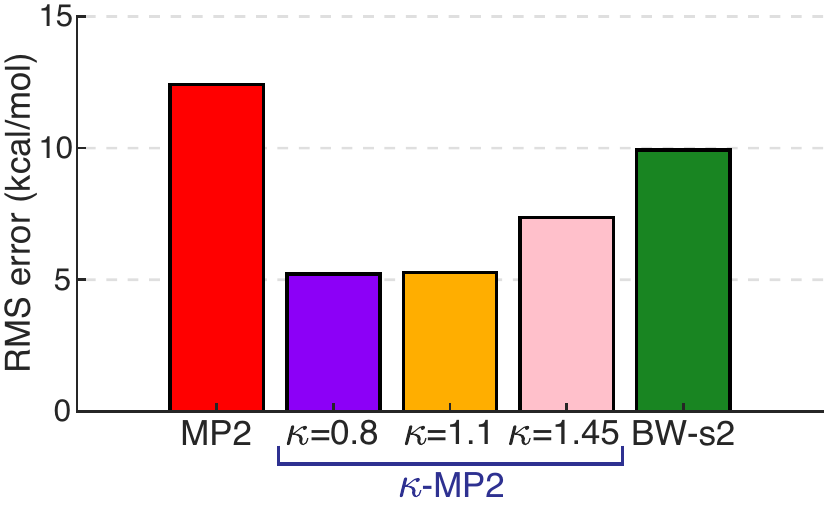}
\caption{
	Root-mean-squared error for the MOR41 dataset
	for MP2, $\kappa$-MP2, and BW-s2 using the
	def2-TZVPP basis along with the def2-ECP for
	4d and 5d metal atoms. An optimal value of $\kappa=0.8$
	was determined in Ref.~\citen{SheRoiLet21}.
}\label{fig:MOR39}
\end{figure}

Another setting where regularized MP2 seems to thrive whereas
MP2 often fails is in transition metal systems.\cite{SheRoiLet21}
To assess our approach, we report
a finite-basis set comparison
of MP2, $\kappa$-MP2, and BW-s2 against the
def2-TZVPP data of the metal-organic reactions (MOR41) dataset
in Fig.~\ref{fig:MOR39}.\cite{DohHanSte18}
For transition metal systems, MP2 performs poorly with
an RMSE of 12.4~kcal\slash mol
and $\kappa$-MP2 with a very low $\kappa=0.8$
performs quite admirably with an RMSE of 5~kcal\slash mol.
However, this value of $\kappa$ represents very strong regularization,
and values of $\kappa=1.1$ or $\kappa=1.45$ are more
appropriate for general usage. The error does not increase when going
to $\kappa=1.1$, but it increases noticeably to 7.4~kcal\slash mol
with $\kappa=1.45$.
While BW-s2 does not perform as well as $\kappa$-MP2
on MOR41, the RMSE is still reduced by
2.4~kcal\slash mol relative to MP2.

Regarding the $\kappa$-MP2 results,
the parameter $\kappa=0.8$ is very small
and performs poorly for NCIs, barrier heights,
and thermochemical properties,
suggesting that it is highly adapted to transition metal complexes.
While the results in Fig.~\ref{fig:MOR39}
highlight some limitations in the flexibility
of BW-s2, whose errors are most similar to $\kappa$-MP2($\kappa=1.45$),
it also shows that empirical parameterization can
be tailor made for a given class of chemical problem.
Overall, BW-s2 provides a satisfactory improvement relative
to MP2 for transition metal systems
while remaining comparable to $\kappa$-MP2 within a more typical
$\kappa$ parameter range.

\section{Conclusions}
We have suggested a
novel partitioning of the Hamiltonian into
a zero-order part that includes the usual sum of
Fock operators along with a regularizer operator
that acts to screen the pair correlations in the resultant theory.
We cast the second-order Brillouin Wigner energy from this theory
into a tensor framework such that orbital invariance
was straightforwardly preserved,
and we chose a form
of the regularizer operator that resulted in a
size-consistent and size-extensive second-order energy.
We also suggested a general algorithm to solve
the self-consistent second-order equation
at ${\mathcal O}(N^5)$ cost and without the need to store
amplitudes.


Over a small set of single, double, and triple bond dissociations, second-order size-consistent Brillouin-Wigner
perturbation theory with a shifted zero-order Hamiltonian
(BW-s2) performs encouragingly by dissociating the C--C bond
in ethane to an asymptotic limit that is invariant to the
spin-polarization of the reference orbitals, while supplying
smooth potential energy curves for multiple-bond
dissociation in ethene and nitrogen. Our approach
is exact for two-electron, two-orbital systems
and dissociates minimum basis H$_2$ to the full configuration interaction
limit regardless of the choice of reference orbitals.
The BW-s2 approach also
performs about as well as the $\kappa$-MP2
approach across noncovalent interactions of small molecules,
and while performing only slightly worse than $\kappa$-MP2($\kappa=1.45$)
for metal\slash organic reaction
barriers and noncovalent interaction energies of nanoscale
$\pi$-stacked systems, BW-s2 still
improves significantly upon the MP2 results.
Importantly, for
broad thermochemical properties BW-s2 outperforms $\kappa$-MP2
by a wide margin, even nearing the performance of CCSD.

The L7 and MOR41 datasets require
exceptionally strong regularization for $\kappa$-MP2
to be successful ($\kappa=1.1$ and $\kappa=0.8$, respectively).
In these cases, BW-s2 does not match the accuracy
of $\kappa$-MP2 as it generally supplies
softer regularization that tends to be more comparable to
a more conservative $\kappa$-MP2($\kappa=1.45$).
So, while BW-s2 is less flexible than empirically
parameterized regularizers, it still gives results that are consistent
with typical values of $\kappa$ when the requisite $\kappa$-regularizer
becomes extreme.
All of this was accomplished with an {\em ab initio}
partitioning of the Hamiltonian, which itself could be
parameterized to augment the strength of the regularization.


\begin{acknowledgments}
This work was supported by the Director, Office of Science,
Office of Basic Energy Sciences,
of the U.S. Department of Energy under Contract No. DE-AC02-05CH11231.
K. C.-F. acknowledges support from the National Institute Of
General Medical Sciences of the National Institutes of Health
under Award Number F32GM149165.
The content is solely the responsibility of the authors and does not
necessarily represent the official views of the National Institutes of Health.
\end{acknowledgments}

\section*{Author Declarations}
\subsection*{Conflict of Interest}
Martin Head-Gordon is a part-owner of Q-Chem, which is
the software platform used to perform the developments
and calculations described in this work.

\subsection*{Author Contributions}
\textbf{Kevin Carter-Fenk}: Conceptualization (equal); investigation (equal); writing -- original draft (lead); formal analysis (lead); writing -- review and editing (equal); Software (lead); funding acquisition (equal).
\textbf{Martin Head-Gordon}: Conceptualization (equal); investigation (equal); writing -- review and editing (equal); funding acquisition (equal); Supervision (lead).

\section*{Data Availability}
Cartesian coordinates for each point along the bond-dissociation
potential energy curves
are available in the supplementary material.
Any other data that support this study are available
from either corresponding author upon reasonable request.

\section*{Appendix}
\subsection{Additional notes on the form of \textbf{W}}
Let us consider the simple case of a system with a single doubly-occupied
orbital and $n_v$ virtual orbitals.
In this case, Eq.~\ref{eq:MFCtensor} can be iteratively solved directly
in the canonical molecular orbital basis, as \textbf{W} is trivially diagonal
when only one occupied orbital is present
({\em i.e.} the orbitals are fixed in the canonical representation).
The amplitudes take the form $t_{ii}^{ab}$, and the matrix elements
$W_{ii}$ work out to be,
\begin{equation}
	W_{ii} = \sum\limits_{ab}t_{ii}^{ab}(ii||ab) = 2E_c^{\text{IP},i} \;.
\end{equation}
When we consider constructing the full denominator with $\Delta_{ii}^{ab}$
and $R_{ii}^{ab}$ (Eq.~\ref{eq:Rtensor}), it becomes apparent that
this factor of two is necessary,
\begin{equation}
	R_{ii}^{ab} = \frac{1}{2}(W_{ii}+W_{ii}) = 2E_c^{\text{IP},i} \; ,
\end{equation}
leading to the full denominator,
\begin{equation}
\begin{split}
	\Delta_{ii}^{ab} + R_{ii}^{ab} &= \varepsilon_a + \varepsilon_b - 2\varepsilon_i + 2E_c^{\text{IP},i}\\
	&\qquad= 2E^{\text{IP},i} + 2E_c^{\text{IP},i} - E^{\text{EA},a} - E^{\text{EA},b}
\end{split}
\end{equation}
Therefore the factor of 2$\times$ the ionization energy
is required to augment both
occupied orbital energies (ionization potentials)
by the correlation contribution.

We note that analysis of the MP2 correlation energy in terms of
Koopmans' theorem has been employed to understand
why MP2 energy denominators are typically overestimated,
even in manifestly nondegenerate cases; namely that this can be
understood in terms of missing
particle-hole interactions which would otherwise stabilize
the zeroth-order double-excitations.\cite{Fin16}
The contribution of $E^{\text{IP,corr}}_i$ is
consistently positive, though for any given orbital pair,
the corresponding elements
$W_{ij}$ are not necessarily negative.
The overall effect of this is to increase the energy gaps, but $\mathbf{W}$
still incorporates
off-diagonal contributions that destabilize the final orbital energies.

\subsection{Proof of BW-s2 Size-Consistency}
We consider two closed shell subsystems, $A$ and $B$, that
are infinitely far apart. As the subsystems are isolated
from one another and the BW-s2 energy is orbital invariant
by nature of the tensor formulation, we can cleanly
ascribe occupied and virtual orbitals to each subsystem.
We first examine the form of the $\mathbf{W}$ matrix, Eq.~\ref{eq:Wmatrix}, in this localized orbital basis,
\begin{equation}
    \mathbf{W} =
    \begin{bmatrix}
        \mathbf{W}_{AA} & \mathbf{W}_{AB}\\
        \mathbf{W}_{BA} & \mathbf{W}_{BB}
    \end{bmatrix}
\end{equation}
where $A$ and $B$ denote the subsystem.
In this form, we can readily rule out the
cross terms by examining,
\begin{equation}
    W_{i_Aj_B} = \frac{1}{2}
    \sum\limits_{PQR}
    \sum\limits_{k_Pa_Qb_R}t_{i_Ak_P}^{a_Qb_R}(j_Bk_P||a_Qb_R)+t_{j_Bk_P}^{a_Qb_R}(i_Ak_P||a_Qb_R)
\end{equation}
where $P$, $Q$, and $R$ run over $A$ and $B$ subsystem indexes.
In the case $P=A$, the integrals $(j_Bk_A||ab)=0$,
which includes $t_{j_Bk_A}^{ab}$,
and in the case $P=B$, all integrals $(i_Ak_B||ab)=0$,
resulting in $W_{i_Aj_B}=0~\forall~k_P$.
The only terms that survive are those where $i$, $j$
and $k$ belong to the same subsystem. Of those,
the integrals $(i_Ak_A||a_Bb_A)=(i_Ak_A||a_Ab_B)=(i_Ak_A||a_Bb_B)=0$ as these are excitations from
occupied orbitals in one subsystem to virtual orbitals
in another. Thus, the only integrals that are nonzero
are those that satisfy $\{i, j, k, a, b\}\in A$
or $\{i, j, k, a, b\}\in B$.
The matrix $\mathbf{W}$ therefore takes the form,
\begin{equation}
    \mathbf{W} = 
    \begin{bmatrix}
        \mathbf{W}_{AA} & \mathbf{0}\\
        \mathbf{0} & \mathbf{W}_{BB}
    \end{bmatrix}
\end{equation}
whenever $i$ and $j$ are disjoint. 
This result verifies that $\mathbf{W}$ itself does not couple non-interacting subsystems, just like the Fock matrix, $\mathbf{F}$ or the two-electron integral tensor. In addition, just like $\mathbf{F}_{AA}$, contributions to $\mathbf{W}_{AA}$ are completely independent of the presence of $B$ and vice-versa.

Next, we turn to the correlation energy,
which can be written in the dressed orbital basis as,
\begin{equation}
    E_c^{(2)} = -\frac{1}{4}\sum\limits_{ijab} \frac{|(ij||ab)|^2}{\Delta_{ij}^{ab}+\frac{1}{2}(W_{ii}+W_{jj})}
\end{equation}
Since we ruled out any cross terms from $\mathbf{W}$
above, $W_{ii}$ and $W_{jj}$ simply shift the orbital
energies $i$ and $j$ within their respective subsystems, independent of the presence of other subsystems.
This establishes the BW-s2 energy of each subsystem is independent of the other, since
we can tag each molecular orbital
with a subsystem index and repeat the above exercise
that was carried out for the elements of $\mathbf{W}$,
\begin{equation}
    E_c^{(2)} = -\frac{1}{4}
    \sum\limits_{PQRS}
    \sum\limits_{i_Pj_Qa_Rb_S} \frac{|(i_Pj_Q||a_Rb_S)|^2}{\Delta_{i_Pj_Q}^{a_Rb_S}+\frac{1}{2}(W_{i_Pi_P}+W_{j_Qj_Q})}
\end{equation}
Hence, we find that the only terms that survive
are
$\{i, j, a, b\}\in A$ and $\{i, j, a, b\}\in B$.
Thus, BW-s2 is size-consistent and by trivial
extension, size-extensive.

\section*{References}

%

\end{document}